\documentstyle[aps,prl,epsf,twocolumn,cite,amsmath]{revtex}

\begin{document}
\draft
\title{Field-Induced Magnetic Ordering in the Quantum Spin System KCuCl$_3$}

\author{A. Oosawa,$^1$ T. Takamasu,$^2$ K. Tatani,$^3$ H. Abe,$^2$ N. Tsujii,$^2$
O. Suzuki,$^2$ H. Tanaka,$^1$ G. Kido$^2$ and K. Kindo,$^{3,4}$}

\address{
$^1$Department of Physics, Tokyo Institute of Technology, 
Oh-okayama, Meguro-ku, Tokyo 152-8551, Japan \\
$^2$Department of Nanomaterials Physics, National Institute for Materials Science, 
Sakura, Tsukuba, Ibaraki 305-0003, Japan \\
$^3$Research Center for Materials Science at Extreme Conditions (KYOKUGEN), Osaka University, 
Toyonaka, Osaka 560-0043, Japan \\
$^4$CREST, Japan Science and Technology, Kawaguchi, Saitama 332-0012, Japan
}

\date{\today}

\maketitle

\begin{abstract}
KCuCl$_3$ is a three-dimensional coupled spin-dimer system and has a singlet ground state with an excitation gap ${\Delta}/k_{\rm B}=31$ K. High-field magnetization measurements for KCuCl$_3$ have been performed in static magnetic fields of up to 30 T and in pulsed magnetic fields of up to 60 T. The entire magnetization curve including the saturation region was obtained at $T=1.3$ K. From the analysis of the magnetization curve, it was found that the exchange parameters determined from the dispersion relations of the magnetic excitations should be reduced, which suggests the importance of the renormalization effect in the magnetic excitations. The field-induced magnetic ordering accompanied by the cusplike minimum of the magnetization was observed as in the isomorphous compound TlCuCl$_3$. The phase boundary was almost independent of the field direction, and is represented by the power law. These results are consistent with the magnon Bose-Einstein condensation picture for field-induced magnetic ordering.
\end{abstract}


\section{Introduction}

Recently, the physics of coupled antiferromagnetic spin-dimer systems has been attracting considerable attention from the viewpoints of magnetic excitations \cite{Garrett,Kato1,Cavadini1,Kageyama1,Totsuka1,Xu,Mueller,Cavadini2,Oosawa1,Stone}, the magnetization processes including magnetization plateaus \cite{Narumi,Shiramura1,Totsuka2,Kolezhuk,Kageyama2,Miyahara,Onizuka,Uchida1} and field-induced three-dimensional (3D) magnetic ordering\cite{Giamarchi,Nikuni,Wessel1,Wessel2,Oosawa2,Oosawa3,Tanaka1}. This study is concerned with the magnetization process of and the field-induced 3D magnetic ordering in KCuCl$_3$, which is an $S=\frac{1}{2}$ coupled antiferromagnetic spin-dimer system and has a singlet ground state with an excitation gap (spin gap) ${\Delta}/k_{\rm B}=31$ K \cite{Kato1,Cavadini1,Tanaka2,Shiramura2,Tanaka3}. 

KCuCl$_3$ has a monoclinic structure (space group $P2_1/c$) \cite{Willett}. The crystal structure is composed of planar dimers of Cu$_2$Cl$_6$. The dimers are stacked on top of one another to form infinite double chains parallel to the crystallographic $a$-axis. These double chains are located at the corners and center of the unit cell in the $b-c$ plane as shown in Fig. 1. From the structural point of view, KCuCl$_3$ was first assumed to be a double chain spin system \cite{Tanaka2,Shiramura2}. The magnetic excitations in KCuCl$_3$ have been extensively investigated through neutron inelastic scattering \cite{Kato1,Cavadini1,Kato2,Cavadini3,Cavadini4,Cavadini5,Kato3} and ESR measurements \cite{Tanaka3}. The effective dimer approximation \cite{Cavadini1,Kato3} and the cluster series expansion \cite{Mueller} were successfully applied to analyze the dispersion relations obtained. From these analyses, it was found that the origin of the spin gap is the strong antiferromagnetic interaction in the chemical dimer Cu$_2$Cl$_6$, and that the neighboring dimers are weakly coupled not only along the double chain, but also in the $(1, 0, -2)$ plane, in which the hole orbitals of Cu$^{2+}$ spread. Consequently, KCuCl$_3$ was characterized as a weakly and three-dimensionally coupled spin-dimer system.  

KCuCl$_3$ differs from isostructural TlCuCl$_3$ and NH$_4$CuCl$_3$ in magnetic character. TlCuCl$_3$ also has a gapped ground state, however, the gap is considerably suppressed due to strong interdimer interactions \cite{Cavadini2,Oosawa2,Tanaka1,Shiramura2,Takatsu}. On the other hand, NH$_4$CuCl$_3$ has a gapless ground state at zero magnetic field \cite{Budhy1}. The remarkable feature of NH$_4$CuCl$_3$ is the existence of magnetization plateaus at one-quarter and three-quarters of the saturation magnetization \cite{Shiramura1,Budhy2}.

When a magnetic field is applied in the spin gap system, the gap $\Delta$ is suppressed and closes at $H_{\rm g}=\Delta/g\mu_{\rm B}$. For $H>H_{\rm g}$ the system can undergo magnetic ordering due to 3D interactions with decreasing temperature. Such field-induced magnetic ordering was observed in the isostructural TlCuCl$_3$ \cite{Oosawa2,Oosawa3,Tanaka1}. For the field-induced magnetic ordering, two characteristic features which cannot be explained by the mean-field approach from the real space \cite{Tachiki1,Tachiki2} have been observed, irrespective of the applied field direction, when the applied field $H$ is slightly higher than $H_{\rm g}$. One is that the magnetization has a cusplike minimum at the transition temperature $T_{\rm N}$. The other is that the phase boundary between the paramagnetic phase and the ordered phase can be expressed by the power law 
\begin{equation}
[H_{\rm N}(T)-H_{\rm g}] \propto T^{\phi},
\end{equation}
with $\phi=2.1(1)$. Here, $H_{\rm N}(T)$ denotes the transition field at temperature $T$. These features can be understood in terms of the Bose-Einstein condensation (BEC) of excited triplets (magnons) \cite{Giamarchi,Nikuni,Wessel1,Wessel2}.

When $H$ is slightly higher than $H_{\rm g}$, the density of magnon $n$ which corresponds to the uniform magnetization is small. In this case, the system can be mapped onto the dilute magnon model. Using the Hartree-Fock approximation, Nikuni {\it et al.} \cite{Nikuni} demonstrated that the magnetization has the cusplike minimum at the transition temperature where the BEC of magnons occurs and that the phase boundary can be described by eq. (1) with $\phi=\frac{3}{2}$. Although the theoretical exponent is somewhat smaller than the experimental value $\phi=2.1$, the magnon BEC theory gives a good description of the experimental results for TlCuCl$_3$.

The magnetic anisotropy is negligible in KCuCl$_3$ \cite{Shiramura2} as it is in TlCuCl$_3$ \cite{Oosawa2,Oosawa3}. Therefore, the previously mentioned characteristic features for the magnon BEC should also be observed in KCuCl$_3$, because the magnon BEC is universal in the isotropic spin gap system. With this motivation, we carried out magnetization measurements in static high magnetic fields above the gap field $H_{\rm g} \approx 20$ T using a hybrid magnet. Since the entire magnetization curve in KCuCl$_3$ has not been observed, we also performed magnetization measurements in pulsed high magnetic fields of up to 60 T. As shown below, the saturation of the magnetization was observed at $H_{\rm s}\sim 50$ T. From the saturation field $H_{\rm s}$, it was found that the exchange parameters determined from the analyses of the dispersion should be reduced. This suggests the importance of the renormalization effect in the magnetic excitations in KCuCl$_3$.

\section{Experimental Details}

The single crystals of KCuCl$_3$ were prepared by the Bridgman method. The details of sample preparation were almost the same as those for TlCuCl$_3$ \cite{Oosawa2}. For KCuCl$_3$, the temperature of the center of the furnace was set at 500$^{\circ}$C. Single crystals of size $0.5{\sim}5$ cm$^3$ were obtained. The crystals are easily cleaved along the $(1, 0, {\bar 2})$ plane. The second cleavage plane is $(0, 1, 0)$. These cleavage planes are perpendicular to each other.

The high-field magnetization process in KCuCl$_3$ was measured at $T=1.3$ K using an induction method with a multilayer pulse magnet at the Research Center for Materials Science at Extreme Conditions (KYOKUGEN), Osaka University. The magnetic fields were applied perpendicularly to the $(1, 0, {\bar 2})$ plane and along the $[2, 0, 1]$ direction which is parallel to both cleavage $(1, 0, {\bar 2})$ and $(0, 1, 0)$ planes. We used a single crystal in the measurement for $H {\parallel} [2, 0, 1]$. For $H {\perp} (1, 0, {\bar 2})$ measurement, we stacked several single crystals along the cylindrical axis of the sample holder.

The temperature dependence of the magnetization of KCuCl$_3$ was measured in static magnetic fields of up to 30 T by a sample-extraction method with a 40 T-class hybrid magnet at the National Institute for Materials Science, Tsukuba. The magnetic fields were applied perpendicularly to the cleavage $(0,1,0)$ and $(1,0,{\bar 2})$ planes, and along the $[2,0,1]$ direction. These field directions were perpendicular to one another.

\section{Results and Discussions}

\subsection{Magnetization Process}

Figure 2 shows the magnetization curves of KCuCl$_3$ measured at $T=1.3$ K for $H\perp (1,0,{\bar 2})$ and $H\parallel [2,0,1]$. Here, the magnetization curves are normalized with the $g$-factors, {\it i.e.}, $g = 2.26$ for $H \perp (1,0,{\bar 2})$ and $g = 2.04$ for $H \parallel [2,0,1]$, which were obtained by ESR measurements. The magnetization curves for two different field directions almost coincide with each other when normalized by the $g$-factor, although the data for $H \parallel [2,0,1]$ are somewhat scattered. This implies that the difference between the transition fields $H_{\rm g}$ and the saturation fields $H_{\rm s}$ for $H \perp (1,0,{\bar 2})$ and $H \parallel [2,0,1]$ is attributed to the anisotropy of the $g$-factor, and that the magnetic anisotropy is negligible, as previously concluded \cite{Shiramura2}. Due to the spin gap, the magnetization is almost zero up to the transition field $(g/2)H_{\rm g}=23$ T, and increases rapidly and monotonically, and then saturates at $(g/2)H_{\rm s}=54$ T. As shown in Fig. 2, the slope of the magnetization in the vicinity of $H_{\rm g}$ and $H_{\rm s}$ is steeper than that in the intermediate field region. This behavior arises due to the quantum fluctuation. 

In the magnetization slope region between $H_{\rm g}$ and $H_{\rm s}$, transverse spin components have a long-range order. Since the lowest magnetic excitations occur at ${\it{\pmb Q}}=(0, 0, 1)$ and its equivalent reciprocal lattice points, we assume that the spin ordering equivalent to that observed in TlCuCl$_3$ is realized in the ordered state (see Fig. 3). For the exchange interactions shown in Fig. 3, we use the notation given in ref. \cite{Mueller}: the main intradimer exchange is denoted
as $J$. $J_{(lmn)}$ and $J'_{(lmn)}$ denote the exchange interactions between dimers separated by a lattice vector $l{\it{\pmb a}} + m{\it{\pmb b}} + n{\it{\pmb c}}$. Applying the mean-field approximation to the interdimer interactions, we calculate the magnetization curve for the ground state. The details of the calculation are given in the Appendix. 

In Table I, we list the exchange parameters in KCuCl$_3$ as determined by M\"{u}ller and Mikeska \cite{Mueller}. They applied a cluster series expansion to analyze the dispersion relations observed in KCuCl$_3$ \cite{Cavadini1}. Their theory describes the experimental results, and the individual interdimer exchange parameters which cannot be obtained by the effective dimer approximation were determined. However, they used the same intradimer exchange interaction, $J=4.25$ meV, as that obtained from the effective dimer approximation \cite{Cavadini1}.

The dotted line in Fig. 4 is the magnetization curve calculated with the exchange parameters obtained by M\"{u}ller and Mikeska \cite{Mueller}. We also plotted the experimental result obtained at $T=1.3$ K for $H\perp (1, 0, {\bar 2})$. The calculated gap field $H_{\rm g}$ coincides with the experimental value, while the calculated saturation field $H_{\rm s}$ is considerably larger than the experimental value. We infer that the saturation field of eq. (A9) in the Appendix is close to the rigorous solution, because even for the $S=\frac{1}{2}$ 1D antiferromagnet, the classical calculation gives the same saturation field as the rigorous solution \cite{Griffiths}. Therefore, we have to reduce the exchange parameters. However, it is difficult to adjust all of the exchange parameters, because we have only two equations for $H_{\rm g}$ and $H_{\rm s}$. Thus, we make a uniform reduction for all of the interdimer exchange parameters. The thin solid line in Fig. 4 is the result calculated with the parameters listed in Table I. The intradimer interaction $J$ and the interdimer interaction $J_{(lmn)}$ or $J'_{(lmn)}$ are reduced by factors 0.9 and 0.8, respectively. Both the calculated gap field $H_{\rm g}$ and saturation field $H_{\rm s}$ coincide with experimental values. Hence, we suggest that the renormalization effect in the magnetic excitations cannot be neglected, although the present system is 3D. The renormalization factor $\pi/2$ in the $S=\frac{1}{2}$ 1D antiferromagnet is well known \cite{dCP}. The difference between the calculated and experimental results for $H_{\rm g}<H<H_{\rm s}$ is due to the quantum fluctuation, which is not taken into account in the present calculation. 

\subsection{Field-induced magnetic ordering}

As mentioned in the Introduction, the present system can undergo magnetic ordering at a low temperature, when the magnetic field $H$ is higher than the gap field $H_{\rm g}$. Figure 5 shows the low-temperature magnetization measured at various external fields above $H_{\rm g}$ for $H \parallel b$, $H \perp (1,0,{\bar 2})$ and $H \parallel [2,0,1]$. With decreasing temperature, the magnetization exhibits a cusplike minimum, irrespective of the field direction, as observed in isostructural TlCuCl$_3$. We assign the temperature with a cusplike minimum in the magnetization to the transition temperature $T_{\rm N}$. However, there is a certain degree of error in determining the transition points, because the experimental data are somewhat scattered.

When the magnetic field is slightly higher than the gap field $H_{\rm g}$, the number of created triplets (magnons) is small. Through the transverse components of the interdimer exchange interactions, the magnons can hop to the neighboring dimers in the same way as particles, which have bosonic natures. The longitudinal component of the the interdimer exchange interaction gives rise to the interaction between magnons. Hence, the system can be mapped onto the interacting dilute boson system, and the phase transition can be described by the formation of the coherent state of dilute magnons, {\it i.e.}, Bose-Einstein condensation (BEC) \cite{Giamarchi,Nikuni,Wessel1}. The cusplike minimum of the temperature variation of the magnetization is characteristic of the BEC of the dilute magnons \cite{Nikuni}, and cannot be described by the mean-field approach from the real space \cite{Tachiki1,Tachiki2}. It is known that the mean-field approximation gives a reasonable description of the physical quantities at $T=0$, while it does not at finite temperatures. The mean-field approach gives the temperature-independent magnetization below $T_{\rm N}$. The density of magnons $n$ corresponds to the magnetization per Cu$^{2+}$ ion $m$, {\it i.e.}, $m=(g/2){\mu_B}n$. The reason that the magnetization increases again below $T_{\rm N}$ is because the increase of the number of condensed magnons is greater than the decrease of the number of thermally excited noncondensed magnons, so that the total number of magnons increases below $T_{\rm N}$. The present experimental result is consistent with the magnon BEC theory of the phase transition.

The increase of the magnetization below $T_{\rm N}$ becomes smaller with increasing magnetic field. At $H=30$ T, the density of magnons $n$ is $n\approx 0.2$. For this large value of $n$, the condition of dilute magnons is no longer satisfied. In the dense magnon region, the hopping of magnon is significantly suppressed, and thus, the mean-field approach from the real space may give a better description of the phase transition.

The transition temperature $T_{\rm N}$ and transition field obtained for $H \parallel b$, $H \perp (1,0,{\bar 2})$ and $H \parallel [2,0,1]$ are plotted in Fig. 6. Since the phase boundary depends on the $g$-factor, we normalize the phase diagram by the $g$-factor. The $g$-factors used are $g = 2.05$ for $H \parallel b$, $g = 2.26$ for $H \perp (1,0,{\bar 2})$ and $g = 2.04$ for $H \parallel [2,0,1]$, which were determined by ESR measurements. Figure 7 shows the phase diagram normalized by the $g$-factor. The phase boundaries for three different field directions almost coincide when normalized by the $g$-factor, although the boundary for $H \perp (1,0,{\bar 2})$ tends to deviate from the others at around $(g/2)H\sim 30$ T. This result may reconfirm that the magnetic anisotropy is negligible in KCuCl$_3$, as in TlCuCl$_3$. 

The phase boundary near the gap field can be described by the power law of eq. (1), as predicted by the magnon BEC theory \cite{Nikuni}. We fit eq. (1) to the data for $T < 3$ K, for which $n < 0.03$. The solid line in Fig. 7 is the fitting with $\phi=2.3(1)$ and $(g/2)H_{\rm g}=22.6(1)$ T. This exponent $\phi$ is close to $\phi=2.1(1)$ observed in TlCuCl$_3$ \cite{Oosawa3,Tanaka1}. The exponent $\phi=2.3$ obtained by the present measurements is somewhat larger than the value $\phi=\frac{3}{2}$ predicted by the magnon BEC theory \cite{Nikuni}. We note that the exponent depends on the temperature region used for fitting. If we use the data up to higher temperatures, we have a larger exponent. This is because the ordered phase exists in a closed area in the magnetic field vs temperature diagram, and thus the entire phase boundary cannot be described by the power law with a single exponent. Since the low-temperature data contribute greatly to the determination of the exponent $\phi$ for the phase boundary near the gap field, the measurements at lower temperatures are needed. 

The increase of the transition field obeying the power law with increasing temperature cannot be described in terms of the mean-field approach from the real space \cite{Tachiki1,Tachiki2} or the temperature dependence of the excitation gap \cite{Cavadini5}. The mean-field result gives the transition field almost independent of temperature at low temperatures. The temperature dependence of the excitation gap was investigated by means of neutron scattering \cite{Cavadini5}. The gap is almost independent of temperature up to approximately 10 K, and then increases significantly.

\section{Conclusion}

We have presented the results of magnetization measurements on the coupled spin-dimer system KCuCl$_3$. The entire magnetization curve including the saturation region was obtained at $T=1.3$ K, using pulsed magnetic fields up to 60 T. Applying mean-field approximation on the interdimer interactions, we analyzed the magnetization curve. It was found that the exchange parameters determined from the dispersion relations of the magnetic excitations should be reduced to fit both the gap field and the saturation field. This suggests the importance of the renormalization effect in the magnetic excitations. 

The field-induced magnetic ordering was investigated in static magnetic fields up to 30 T. As observed in isostructural TlCuCl$_3$, the magnetization exhibits a cusplike minimum at the transition temperature. The phase boundary is almost independent of field direction, and is described by the power law. These features are compatible with the magnon BEC theory. Therefore, we conclude that the field-induced 3D magnetic ordering near the gap field in isotropic spin gap systems such as KCuCl$_3$ and TlCuCl$_3$ is universally represented by the BEC of dilute magnons.

\acknowledgments
This work was supported by the Toray Science Foundation and a Grant-in-Aid for Scientific Research on Priority Areas (B) from the Ministry of Education, Culture, Sports, Science and Technology of Japan. A. O. was supported by the Research Fellowships of the Japan Society for the Promotion of Science for Young Scientists.

\appendix
\section{Derivation of the magnetization curve}
We calculate the magnetization curve at $T=0$, assuming that the triplet excitations are created only on the dimer site. The interdimer interactions are treated by the mean-field approximation. In the magnetization slope region, the transverse spin components form the long-range order. Since the lowest magnetic excitations in KCuCl$_3$ and TlCuCl$_3$ occur at the same reciprocal lattice points, {\it i.e.}, ${\it{\pmb Q}}=(h, 0, l)$ with integer $h$ and odd $l$ in the $a^*-c^*$ plane \cite{Kato1,Cavadini1,Cavadini2,Oosawa1}, we can assume that the spin structure in the ordered phase in KCuCl$_3$ is equivalent to that observed in TlCuCl$_3$ (see Fig. 3) \cite{Tanaka1}. We express the spin state of a dimer with the total spin $S=S_1+S_2$ and the $z$ component $S^z=S^z_1+S^z_2$ as $|S,S^z\rangle$, where subscript numbers 1 and 2 distinguish spins in a dimer. In the ordered phase, the transverse components of two spins on a dimer should be antiparallel and their magnitudes should be the same, {\it i.e.,} $\left<S^{x,y}_1\right> = -\left<S^{x,y}_2\right>$. This indicates that the spin state $|1,0\rangle$ does not contribute to the basis state. Thus, we write the basis state of the $j$-th dimer as 
\begin{equation}
\psi_j=\left|0,0\right> \cos{\theta}+\left(\left|1,1\right> \cos{\varphi}\ {\rm e}^{{\rm i}\phi_j} - \left|1,-1\right> \sin{\varphi}\ {\rm e}^{-{\rm i}\phi_j} \right)\sin{\theta}\ . 
\end{equation}
Angles $\theta$ and $\varphi$ were introduced to satisfy the normalization condition. The phase $\phi_j$ corresponds to the angle between the $x$-direction and the transverse component of the spin on the $j$-th dimer. 

The average values of spin operators are given by
\begin{eqnarray}
\left<S^{z}_{j,1} \right> &=& \left<S^{z}_{j,2}\right> = \frac{1}{2}\sin^2{\theta}\cos{2\varphi}, \nonumber \\
\left<S^+_{j,1} \right> &=& -\left<S^+_{j,2} \right>=-\frac{1}{\sqrt{2}} \cos{\theta}\sin{\theta}\left(\cos{\varphi}+\sin{\varphi}\right){\rm e}^{-{\rm i}\phi_j}, \nonumber\\
\left<S^-_{j,1} \right> &=& -\left<S^-_{j,2} \right>=-\frac{1}{\sqrt{2}} \cos{\theta}\sin{\theta}\left(\cos{\varphi}+\sin{\varphi}\right){\rm e}^{{\rm i}\phi_j}\ . 
\end{eqnarray}
From the spin structure shown in Fig. 3, we see that all of the phases $\phi_j$ of dimers are the same in a chemical double chain, and that the phases of dimers in the double chains located at the corner and the center of the unit cell in the $b-c$ plane differ by $\pi$. With eq. (A2) and the exchange interactions shown in Fig. 3, the energy per dimer is expressed as   
\begin{eqnarray} 
E & = & -\frac{3}{4}J\cos^2{\theta}+\frac{1}{4}J\sin^2{\theta}-g\mu_{\rm B}H\sin^2{\theta}\cos{2\varphi} \nonumber \\
&+& {\tilde J}\sin^2{\theta}\cos^2{\theta}\left(1+\sin{2\varphi}\right)+\frac{1}{2}{\bar J}\sin^4{\theta}\cos^2{2\varphi}\ ,
\end{eqnarray}
where
\begin{equation}
{\tilde J} = J^{\rm eff}_{(100)} - 2J^{\rm eff}_{\left(1,\frac{1}{2},\frac{1}{2}\right)} + J^{\rm eff}_{(2,0,1)}\ ,  
\end{equation}
with
\begin{eqnarray}
J^{\rm eff}_{(100)} &=& \frac{1}{2}\left(2J_{(100)} - J'_{(100)}\right)\ ,\nonumber \\
J^{\rm eff}_{\left(1,\frac{1}{2},\frac{1}{2}\right)} &=& \frac{1}{2}\left(J_{\left(1,\frac{1}{2},\frac{1}{2}\right)} - J'_{\left(1,\frac{1}{2},\frac{1}{2}\right)}\right)\ ,\nonumber \\
J^{\rm eff}_{(2,0,1)} &=& - \frac{1}{2}J'_{(2,0,1)}
\end{eqnarray}
and
\begin{equation}
{\bar J} = J_{(100)} + \frac{1}{2}J'_{(100)} + J_{\left(1,\frac{1}{2},\frac{1}{2}\right)} + J'_{\left(1,\frac{1}{2},\frac{1}{2}\right)} + \frac{1}{2}J'_{(2,0,1)}\ .
\end{equation}
Angles $\theta$ and $\varphi$ can be determined by $\partial E/\partial \theta=0$ and $\partial E/\partial \varphi=0$, which lead to
\begin{equation}
J - g\mu_{\rm B}H\cos{2\varphi} + {\tilde J}\cos{2\theta}\left(1+\sin{2\varphi}\right) + {\bar J}\sin^2{\theta}\cos^2{2\varphi} = 0\ , 
\end{equation}
and
\begin{equation}
g\mu_{\rm B}H\sin{2\varphi} + {\tilde J}\cos^2{\theta}\cos{2\varphi} - {\bar J}\sin^2{\theta}\cos{2\varphi}\sin{2\varphi} = 0\ ,
\end{equation}
for the ordered state between the gap field $H_{\rm g}$ and the saturation field $H_{\rm s}$. The magnetization curve is obtained by solving eqs. ({\rm A}7) and ({\rm A}8) self-consistently. 

At the saturation field $H_{\rm s}$, $\sin{\theta}=1$ and $\sin{\varphi}=0$. Thus, $H_{\rm s}$ is given by
\begin{equation}
g\mu_{\rm B}H_{\rm s} = J - {\tilde J} + {\bar J}\ . 
\end{equation}
At the gap field $H_{\rm g}$, $\sin{\theta}=0$. Substituting this condition into eqs. ({\rm A}7) and ({\rm A}8), we obtain 
\begin{equation}
g\mu_{\rm B}H_{\rm g} = \sqrt{J^2 + 2J{\tilde J}}\ .
\end{equation}
It is noted that the right-hand side of eq. ({\rm A}10) is equivalent to the lowest excitation energy given by the effective dimer approximation, in which the individual interdimer interactions $J_{(lmn)}$ and $J'_{(lmn)}$ are reduced to an effective interdimer interaction $J^{\rm eff}_{(lmn)}$ as shown in eq. ({\rm A}5) \cite{Cavadini1}. If we neglect the highest $\left|1,-1\right>$ state, {\it i.e.}, $\sin{\varphi}\equiv 0$, we obtain
\begin{equation}
g\mu_{\rm B}H_{\rm g} = J + {\tilde J}\ .
\end{equation}
This result can also be derived from the Tachiki-Yamada theory \cite{Tachiki1,Tachiki2}.

\newpage

\begin{table}
\caption{Exchange interactions in KCuCl$_3$ in the unit of meV.}
\label{table2}
\begin{tabular}{ccccccc}
$J$ & $J_{(100)}$ & $J'_{(100)}$ & $J_{\left(1,\frac{1}{2},\frac{1}{2}\right)}$ & $J'_{\left(1,\frac{1}{2},\frac{1}{2}\right)}$ & $J'_{(2,0,1)}$ & ref. \\
\tableline
4.25 & $-0.021$ & 0.425 & 0.850 & 0.170 & 0.799 & \cite{Mueller} \\
3.83 & $-0.017$ & 0.340 & 0.680 & 0.136 & 0.639 & this work \\
\end{tabular}
\end{table}

\begin{figure}[ht]
\epsfxsize=70mm
\centerline{\epsfbox{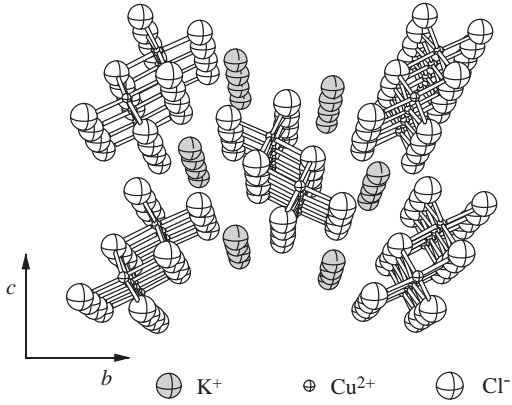}}
\vspace*{1cm}
	\caption{Crystal structure of KCuCl$_3$ viewed along the $a$-axis. Shaded, small open and large open circles denote K$^+$, Cu$^{2+}$ and Cl$^-$ ions, respectively.}
	\label{Fig.1}
\end{figure}


\begin{figure}[ht]
\epsfxsize=70mm
\centerline{\epsfbox{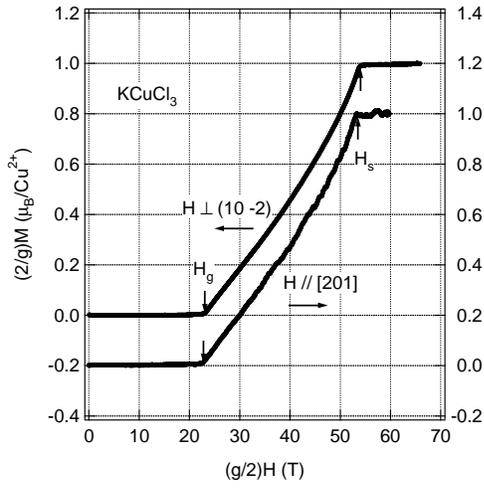}}
\vspace*{1cm}
	\caption{Magnetization curve of KCuCl$_3$ measured at $T=1.3$ K for the magnetic fields $H\parallel [2, 0, 1]$ and $H\perp (1, 0, {\bar 2})$. The values of the magnetization and the magnetic field are normalized by the $g$-factor.}
	\label{Fig.2}
\end{figure}


\begin{figure}[ht]
\epsfxsize=70mm
\centerline{\epsfbox{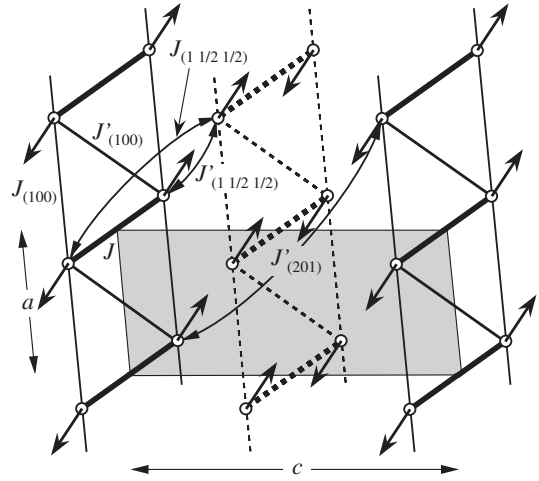}}
\vspace*{1cm}
	\caption{Important exchange interactions in KCuCl$_3$ and the spin structure observed in the ordered phase of TlCuCl$_3$ for $H\parallel b$. The double chain located at the corner and the center of the chemical unit cell in the $b-c$ plane are represented by solid and dashed lines, respectively. The shaded area is the chemical unit cell in the $a-c$ plane.}
	\label{Fig.3}
\end{figure}


\begin{figure}[ht]
\epsfxsize=70mm
\centerline{\epsfbox{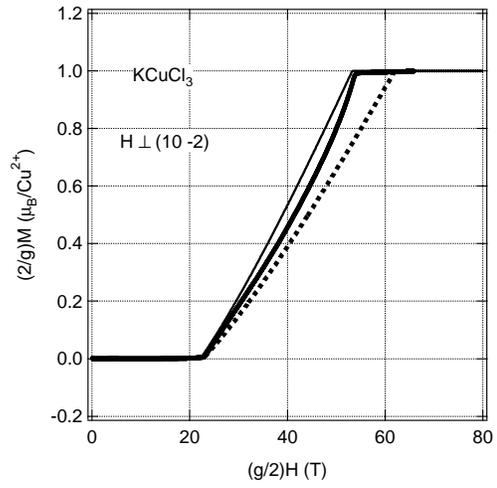}}
\vspace*{1cm}
	\caption{Magnetization curve calculated by the mean-field approximation. The dotted and thin solid lines are the results with the exchange parameters obtained by M\"{u}ller and Mikeska \cite{Mueller}, and reduced ones, respectively. The thick solid line is the magnetization curve observed in KCuCl$_3$ for $H\perp (1, 0, {\bar 2})$. }
	\label{Fig.4}
\end{figure}


\begin{figure}[ht]
\epsfxsize=70mm
\centerline{\epsfbox{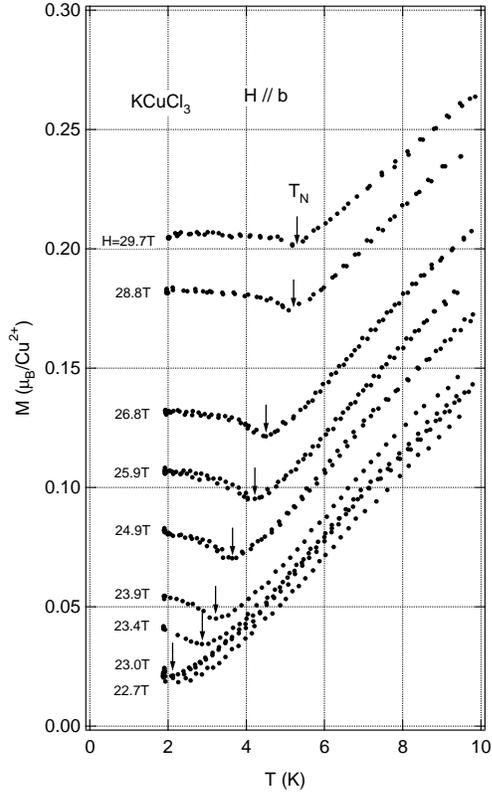}}
\begin{center}
(a)
\end{center}
\epsfxsize=70mm
\centerline{\epsfbox{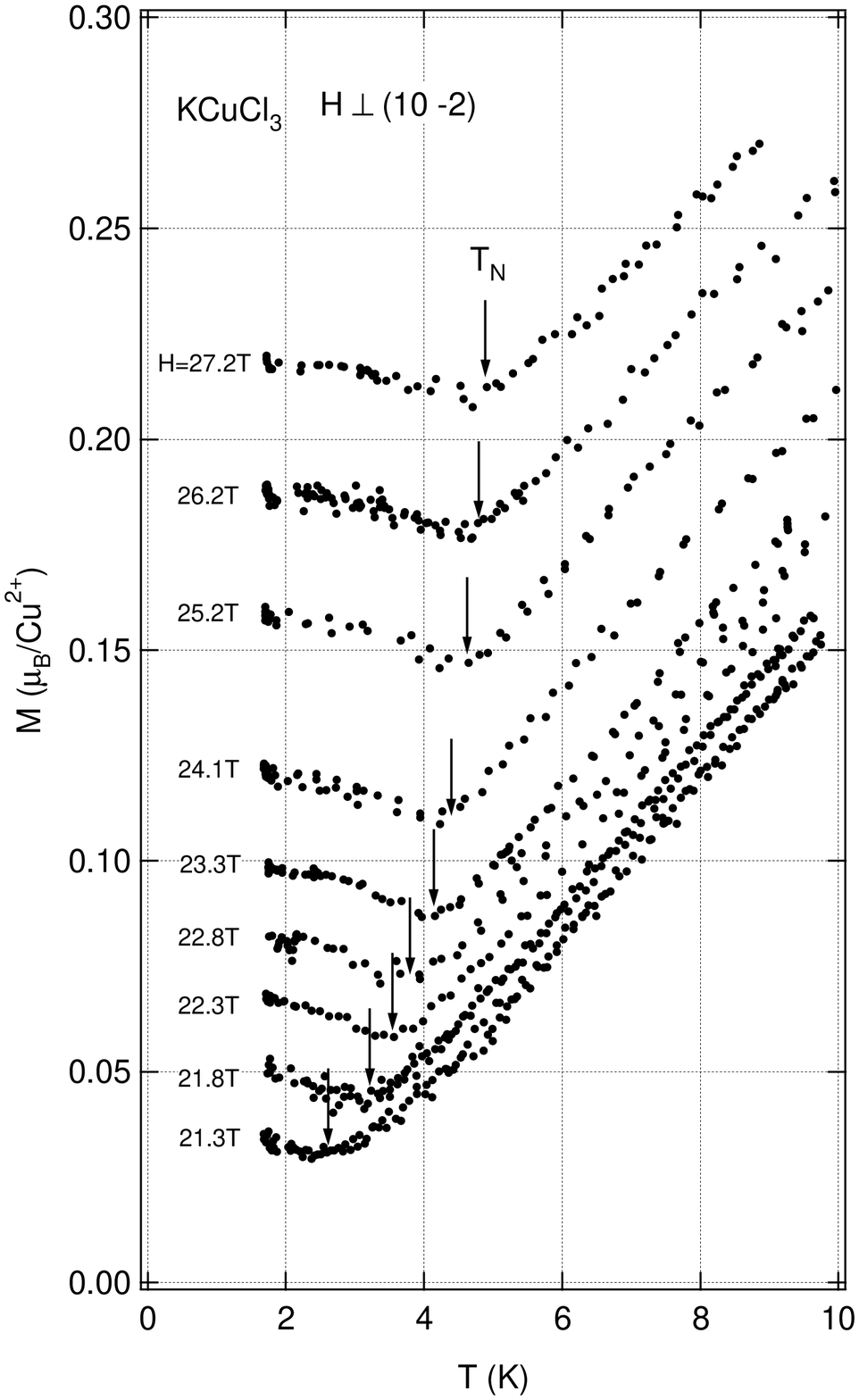}}
\begin{center}
(b)
\end{center}
\epsfxsize=70mm
\centerline{\epsfbox{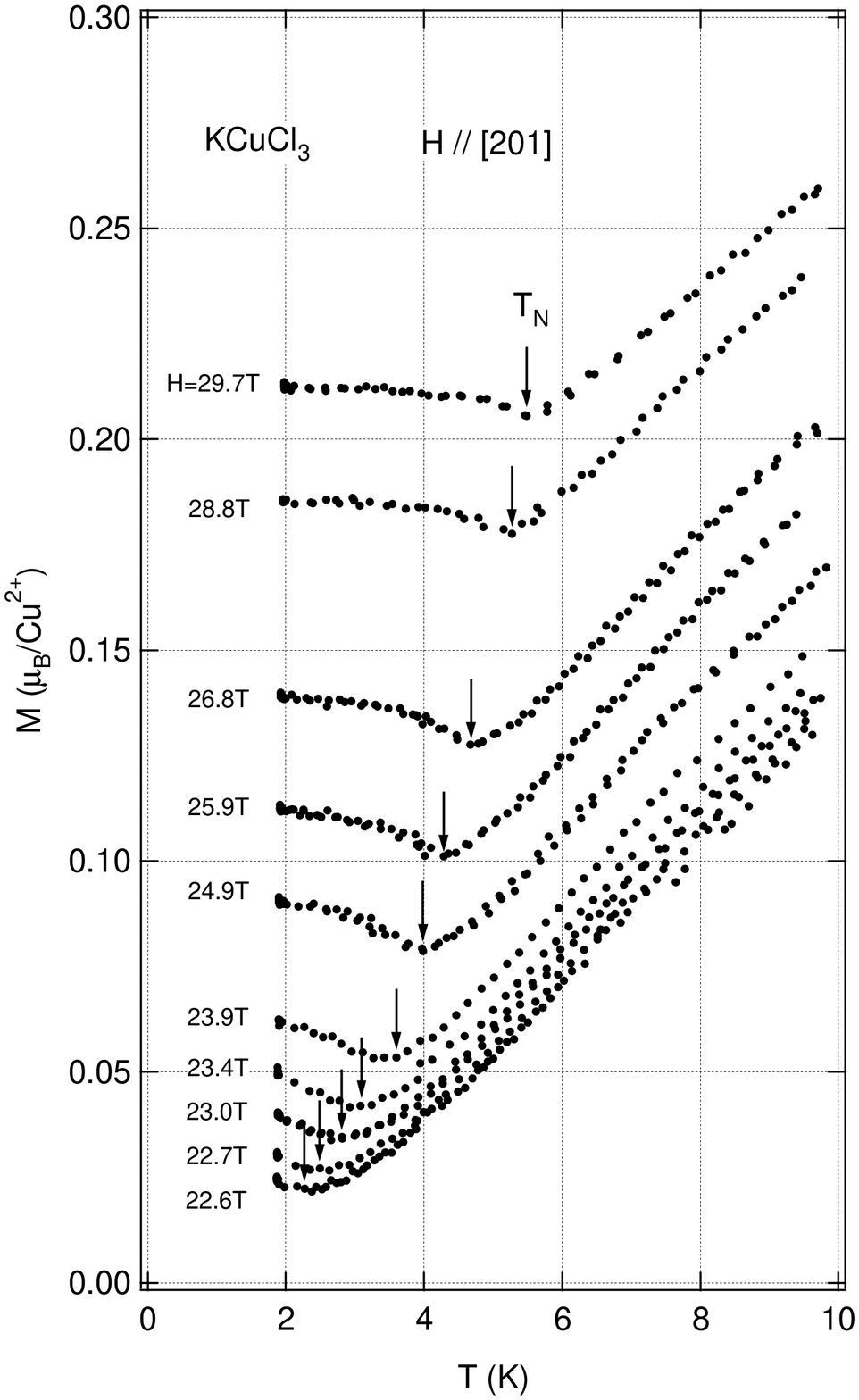}}
\begin{center}
(c)
\end{center}
\vspace*{1cm}
	\caption{Low-temperature magnetizations of KCuCl$_3$ measured at various external fields for (a) $H \parallel b$, (b) $H \perp (1,0,{\bar 2})$ and (c) $H \parallel [2,0,1]$.}
	\label{Fig.5}
\end{figure}


\begin{figure}[ht]
\epsfxsize=70mm
\centerline{\epsfbox{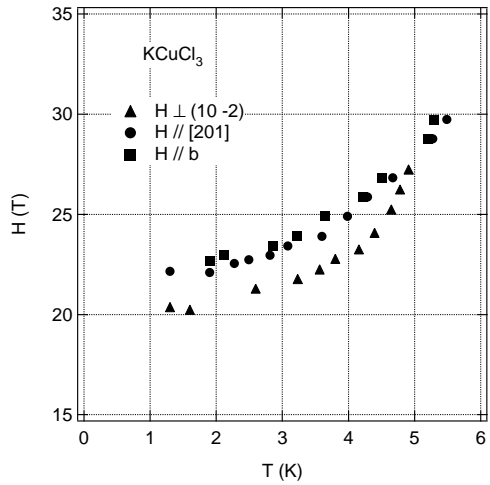}}
\vspace*{1cm}
	\caption{Phase boundaries in KCuCl$_3$ obtained for three different field directions.}
	\label{Fig.6}
\end{figure}


\begin{figure}[ht]
\epsfxsize=70mm
\centerline{\epsfbox{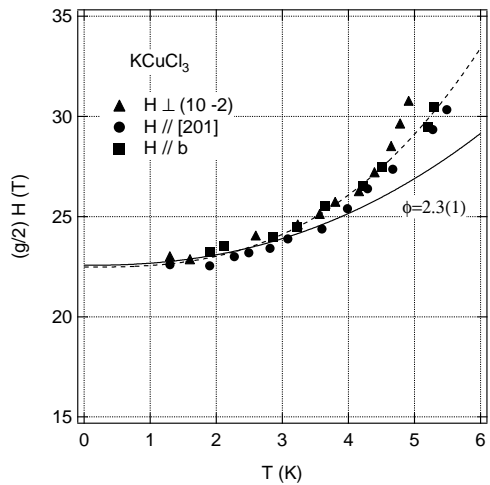}}
\vspace*{1cm}
	\caption{Phase diagram in KCuCl$_{3}$ normalized by the $g$-factor. The solid line denotes the fitting by eq. (1) with $(g/2)H_{\rm g}=22.6(1)$ T and ${\phi}=2.3(1)$. The dashed line is a visual guide.}
	\label{Fig.7}
\end{figure}

\end{document}